\documentclass[aps,prl,twocolumn,showpacs]{revtex4}
\usepackage{upgreek}
\usepackage{graphicx}

\newcommand{\ctr}{}

\begin{document}

\title{NOON states from cavity-enhanced down-conversion:\\High quality and super-resolution}

\author{Florian Wolfgramm, Alessandro Cer\`{e}, and Morgan W. Mitchell}
\address{ICFO - Institut de Ciencies Fotoniques, Mediterranean
Technology Park, 08860 Castelldefels (Barcelona), Spain}

\date{27 November 2009}

\begin{abstract}
Indistinguishable photons play a key role in quantum optical
information technologies.  We characterize the output of an
ultra-bright photon-pair source using multi-particle tomography
[R.~B.~A.~Adamson et al., Phys.~Rev.~Lett.~{\bf 98}, 043601
(2007)] and separately identify coherent errors, decoherence, and
distinguishability.  We demonstrate generation of high-quality
indistinguishable pairs and polarization NOON states with 99\%
fidelity to an ideal NOON state. Using a NOON state we perform a
super-resolving angular measurement with 90\% visibility.
\end{abstract}


\maketitle

\begin{figure}[t]
\centering
\includegraphics[width=7cm]{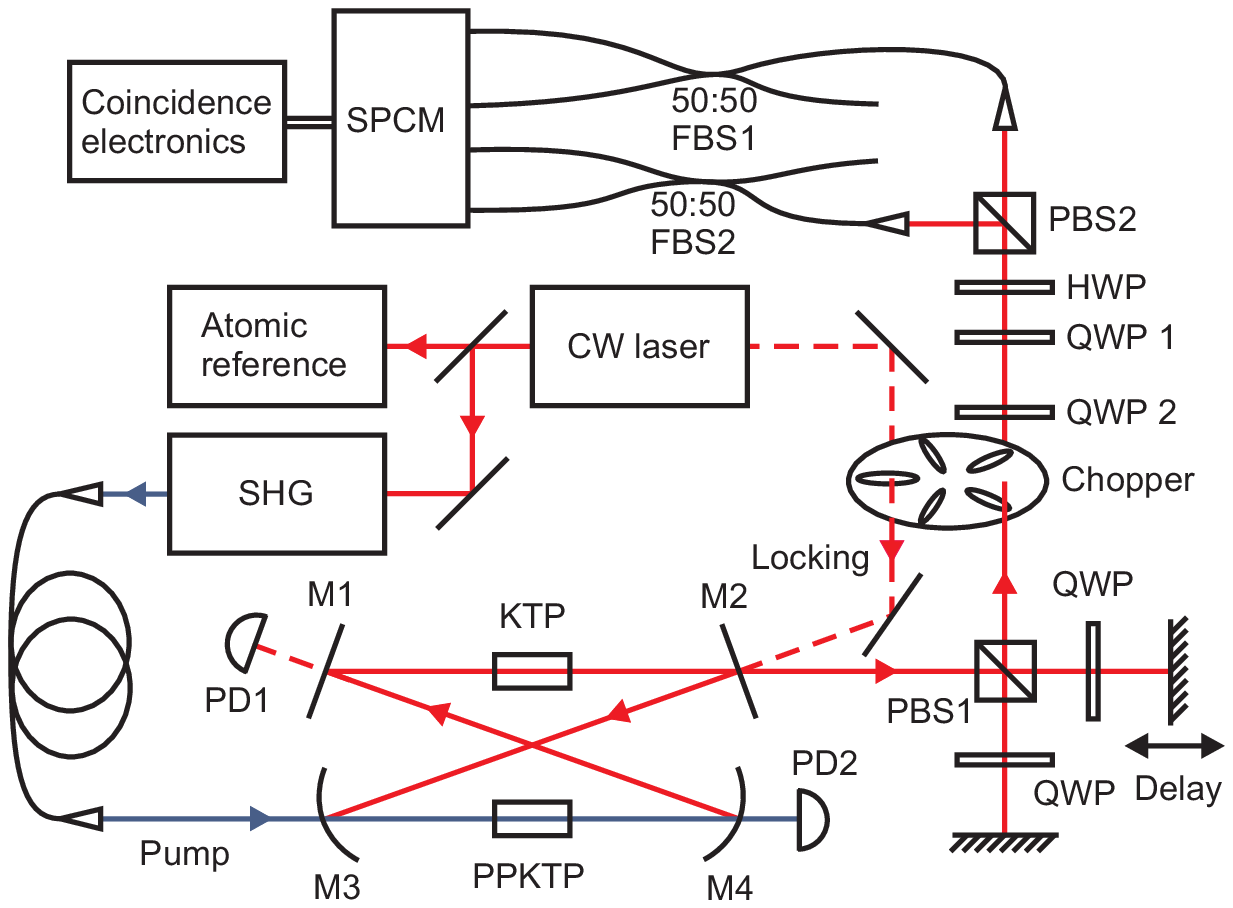}
\caption{Experimental Setup. SHG, second harmonic generation
cavity; PPKTP, phase-matched nonlinear crystal; KTP, compensating
crystal; M1-4, cavity mirrors; PBS, polarizing beam splitter; HWP,
half wave plate; QWP, quarter wave plate; SMF, single-mode fiber;
PD, photodiode; FBS, fiber beam splitter; SPCM, single photon
counting module.\label{img:Setup}}
\end{figure}

\begin{figure}[t]
\centering
\includegraphics[width=6cm]{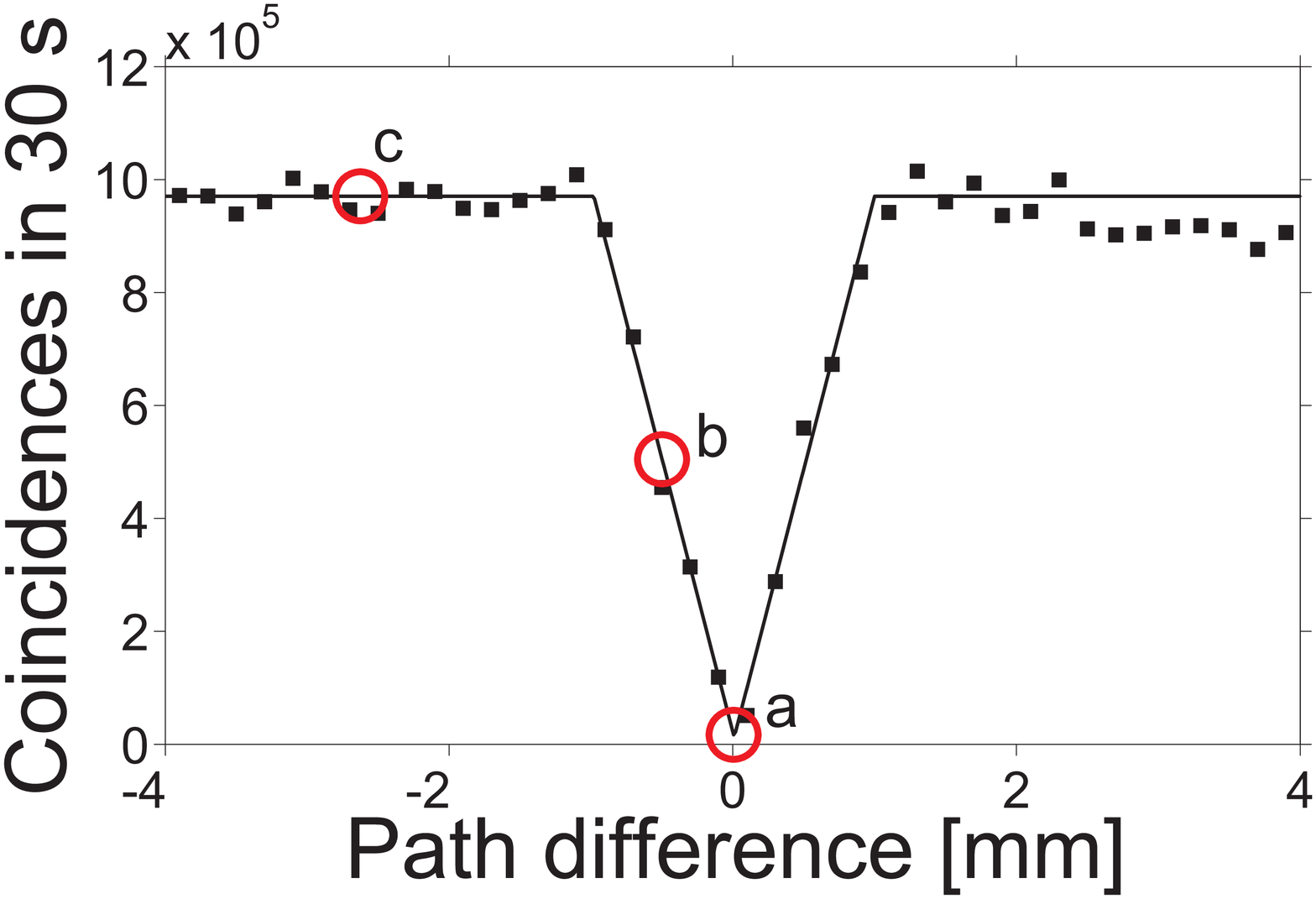}
\caption{Hong-Ou-Mandel effect. Experimental data and triangular
fit function. Labelled points indicate locations of tomographic
reconstructions in
Figure~\ref{img:different_dip_positions}.\label{img:HOM}}
\end{figure}

\begin{figure}[h]
\centering
\includegraphics[width=5cm]{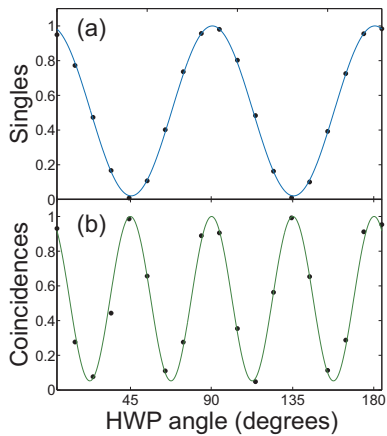}
\caption{(a) Standard phase measurement. Normalized singles
detection at the transmitted port of PBS2. In this measurement
only the $H$ polarized part of the pair-photon state was sent to
the analyzer. (b) Super-resolving phase measurement. Normalized
coincidence detection between reflected and transmitted port of
PBS2 for a NOON state input. The shorter period of the coincidence
counts oscillations indicates
super-resolution.\label{img:Super-Resolution}}
\end{figure}

Many applications in quantum information, quantum imaging and
quantum metrology rely on the availability of high-quality single
photons or entangled photon pairs. Depending on the kind of
application, the requirements on a source of photonic quantum
states include brightness and efficiency as well as the degree of
indistinguishability, purity and entanglement of the output state.
\\
Recently there has been increasing interest in the generation of
cavity-enhanced paired photons \cite{Ou2000, Lu2003, Wang2004,
Kuklewicz2005, Kuklewicz2006, Scholz2007, Wolfgramm2008, Bao2008},
because the cavity geometry enhances the photon generation into
the spatial and spectral resonator modes increasing at the same
time brightness, collection efficiency and indistinguishability.
The enhancement of the brightness is of the order of the cavity
finesse \cite{Kuklewicz2005}. When the nonlinear crystal is
type-II phase-matched, it is possible to achieve polarization
entanglement and create NOON states that can be applied to achieve
phase super-resolution. Due to multiple passage through the
crystal, such cavity schemes are more sensitive to imperfections
in materials and alignment. Another challenge is double resonance
that is not automatically achieved since in a type-II process the
two generated photons have orthogonal polarization and see
different refractive indices in the birefringent nonlinear
crystal. Due to these cavity effects type-II schemes can suffer
from a low Hong-Ou-Mandel (HOM) dip visibility \cite{Hong1987},
e.g. in \cite{Kuklewicz2005}, where the reported visibility was
76.8\% and not all reasons for the low visibility could be
identified. A limited visibility can be caused by distinguishing
timing information, coherent state-preparation errors, and
decoherence. These three possibilities cannot be differentiated by
a HOM measurement. Nevertheless, multi-particle states can be
fully characterized, including decoherence and distinguishability
of particles by tomographic techniques \cite{Adamson2007}. We
apply these techniques to the output pairs from a cavity-enhanced
down-conversion source, and show that cavity-enhanced
down-conversion not only provides a large photon flux, but is also
capable of producing highly indistinguishable photons that can be
used to create interesting and useful quantum states such as a
high-fidelity NOON state.
\\
The experimental setup consists of two parts, one for the
preparation of the state and the other for its analysis. The state
preparation part is based on the high-brightness cavity-enhanced
down-conversion source described in detail in a previous
publication \cite{Wolfgramm2008}. As principal light source we use
a single-frequency diode laser locked to the ${\rm D}_1$
transition of atomic rubidium at 795~nm (Figure\ \ref{img:Setup}).
The frequency doubled part of the laser pumps a type-II
phase-matched PPKTP crystal inside an optical cavity. After the
photons leave the cavity, a variable retarder consisting of a
polarizing beam splitter, two quarter wave plates and two mirrors
in a Michelson geometry produces a relative delay between the $H$-
and $V$-polarized photons.
\\
A general polarization analyzer, consisting of a quarter wave
plate (QWP1) followed by a half wave plate (HWP) and a polarizing
beam splitter (PBS2) is used to determine the measurement basis as
shown in Figure\ \ref{img:Setup}. To generate a NOON state in the
$H$/$V$ basis another quarter wave plate (QWP2) can be added. The
two output ports of PBS2 are coupled to single-mode fibers and
split with 50:50 fiber beam splitters. The four outputs are
connected to a set of single photon counting modules (Perkin Elmer
SPCM-AQ4C). Time-stamping was performed by coincidence electronics
with a resolution of 2~ns. By considering a time window of 150~ns,
that is longer than the coherence time of each individual photon,
we can evaluate the coincidences between any two of the four
channels.
\\
To observe a classic ``HOM dip'', we set the analyzer for $\pm
45$~degree with QWP1 and QWP2 removed and scan the relative delay.
The resulting curve is shown in Figure\ \ref{img:HOM}. Accidental
counts due to probabilistic photon arrival times that account for
approximately 10\% of the coincidence counts out of the dip have
been subtracted.   For zero delay the photons are highly
indistinguishable indicated by a visibility of the graph of 96\%.
With reduced pump power, similar visibilities were seen without
need for accidentals subtraction.
\\
We use the multi-particle state tomography of reference
\cite{Adamson2007}.  As in tomographic characterization of qubits
\cite{James2001}, this method gives a complete description of the
polarization of the photons, including partial coherence.
Moreover, it quantifies the operational distinguishability of the
photons, i.e., information in unobserved degrees of freedom that
could in principle be used to identify the photons, and which
destroys non-classical interference.  For example, photons which
differ in arrival time, frequency, or spatial mode are
operationally distinguishable and do not show non-classical
interference.  The presence of these other degrees of freedom
allows the photons to have any symmetry of their polarization
wave-function while remaining globally symmetric.  Using the
symmetry-ordered basis $\{ \left| H_{1} H_{2} \right>,\left|
\psi^{+} \right>, \left| V_{1} V_{2} \right> , \left| \psi^{-}
\right>\}$, with $\left|  \psi^{\pm} \right> \equiv (\left| H_{1}
V_{2} \right>\pm\left| V_{1} H_{2} \right>/\sqrt{2}$, a general
polarization state is described by a density matrix of the form
\begin{equation}
\rho =
\left(
\begin{array}{cc}
\left(
\begin{array}{ccc}
 & &  \\ & \rho_{S} & \\ & &
\end{array}
\right) &
\begin{array}{c} \cdot \\ \cdot \\ \cdot \end{array}
\\
\begin{array}{ccc} \cdot& \cdot& \cdot \end{array}&
\left(
\begin{array}{c}
 \rho_{A}
\end{array}
\right)
\end{array}
\right),
\end{equation}
where $\rho_S$ is a $3\times 3$ matrix describing the symmetric
portion of the polarization state and $\rho_A$ is a $1\times 1$
matrix describing the anti-symmetric portion.
Coherences~($\cdot$), between these parts are, by assumption,
unobservable and do not contribute to any measurable outcome.
\newcommand{\psp}{\alpha}
\newcommand{\psm}{\beta}
We now calculate the coincidence probability for an arbitrary
state $\rho$, and also for the same state with the photons made
distinguishable.  Applied to the state at the center of a HOM dip,
these give the HOM visibility.  We assume that the analyzer is set
to discriminate in the basis $\left|\psp\right>,\left|\psm\right>
\equiv (\left|H\right> \pm \exp[i \phi] \left|V\right>)/\sqrt{2}$,
so that a coincidence indicates a state with one $\psp$ and one
$\psm$ photon.   Within the symmetric(anti-symmetric) subspace,
the coincidence detection is then a projection onto $(\left|\psp_1
\psm_2\right> \pm\left|\psm_1 \psp_2\right>)/\sqrt{2}$. For the
symmetric subspace this state is
$(\left|H_1,H_2\right>-\exp[2i\phi]
\left|V_1,V_2\right>)/\sqrt{2}$, for the anti-symmetric subspace,
it is $\left|\psi^-\right>$ itself.  In terms of the density
matrix $\rho$, the coincidence probability is then
\begin{equation}
C_{\rm \ctr} = \rho_{44} +\frac{1}{2}( \rho_{11} + \rho_{33}) -
{\rm Re}[e^{2i\phi} \rho_{13}].
\end{equation}
We can also transform this state into one with complete
in-principle distinguishability, by setting
$\rho_{44}=\rho_{22}=(\rho_{22}+\rho_{44})/2$.
Physically, this transformation can be accomplished by introducing
a large delay between the photons.  Using ${\rm Tr}[\rho]=1$ we
find
\begin{equation}
C_{\rm dist} = \frac{1}{2} - {\rm Re}[e^{2i\phi} \rho_{13}].
\end{equation}
If $C_{\rm \ctr}$ is the probability at the center of the HOM dip, i.e.,
for the state with maximal indistinguishability, then the  visibility
of the HOM dip is
\begin{equation}
V_{\rm HOM} \equiv \frac{C_{\rm dist}-C_{\rm \ctr}}{C_{\rm
dist}+C_{\rm \ctr}} =
\frac{\rho_{22}-\rho_{44}}{2-(\rho_{22}-\rho_{44}) - 4{\rm
Re}[e^{2i\phi} \rho_{13}] }.
\end{equation}
We note that this depends on few of the density matrix elements,
and thus a variety of different states could have the same HOM dip
visibility.
\\
We can also calculate the visibility in an interferometric
measurement based on polarization rotations, such as those which
give super-resolution.  We assume a wave plate or other optical
device applies a unitary rotation to both photons of the state,
and they are detected in the $\psp,\psm$ state as above.  The
$\psi^-$ component is invariant under any unitary transformation
affecting both photons, and thus contributes a constant
$\rho_{44}$ to the coincidence probability.  In contrast, the
contribution of the triplet component may oscillate between zero
and \linebreak $\rho_{11}+\rho_{22}+\rho_{33} = 1-\rho_{44}$.  A
limit on interferometric visibility is thus
\begin{equation}
V_{\rm INT} \le \frac{1-\rho_{44}}{1+\rho_{44}}.
\end{equation}
\\
We follow the tomography method developed in \cite{Adamson2007,
Adamson2008} in order to get a polarization characterization of
the output state. We evaluated the coincidence counts for the same
10 different wave plate settings of HWP and QWP1 as in
\cite{Adamson2007}. The acquisition time for each wave plate
setting was 60 seconds. Applying a maximum likelihood
reconstruction we obtain the polarization density matrix. The
single count rate corrected for accidentals during the
measurements was typically 10~000 counts/s.
\\
We measured the density matrix for different delays between the
photons corresponding to different positions in the HOM dip. We
generated different states as follows: a) center of dip b)
mid-point of dip c) outside of dip. In addition, we produced an
unknown state d) by tuning the fundamental laser by about 3.1~GHz
from the frequency used in a) -- c). At this detuning we observe a
reduced HOM visibility, for reasons that are not understood.  Data
for d) were taken at the center of the dip, i.e., with zero
relative delay.
\begin{figure}[b]
\centering
\includegraphics[width=7cm]{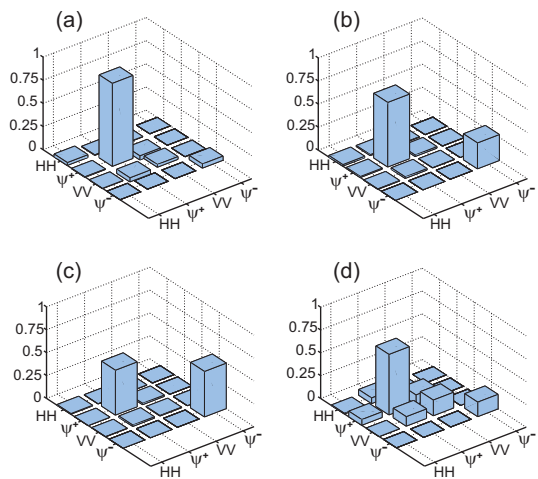}
\caption{Reconstructed polarization density matrices for (a)
center of HOM dip, (b) edge of HOM dip, (c) outside of HOM dip
(corresponding to points in Figure~\ref{img:HOM}), (d) center of
dip, but with system tuned to a different frequency.
\label{img:different_dip_positions}}
\end{figure}
For all these states we applied the same tomography procedure.
Figure~\ref{img:different_dip_positions} shows the elements of the
real parts of the density matrices. The imaginary parts are close
to zero and are not shown.
\\
We note that the populations in $|\psi^+\rangle$ and
$|\psi^-\rangle$ change for different dip positions: a) 94\% and
4\%, b) 68\% and 28\%, c) 49\% and 50\%, d) 66\% and 16\%. We also
note that while b) and d) have a similar amount of HOM dip
visibility, their density matrices look very different. In b) only
the $\psi^+$ and $\psi^-$ populations are significant, while d)
shows also $VV$ population and coherence between $\psi^+$ and
$VV$. Thus b) shows distinguishability while d) shows some
distinguishability but also decoherence and coherent errors which
cause non-zero off-diagonal elements to appear in the density
matrix. This shows clearly that multi-particle tomography provides
information not present in the HOM visibility, and can be useful
for identifying imperfections in generated states.
\\
The achieved high visibility of the state is the requirement for a
high-fidelity NOON state. We introduce another quarter wave plate
(QWP2) before the analyzing part of the setup to create a
two-photon NOON state in the $H$/$V$ basis, which can be written
$1/\sqrt{2}(|H_1,H_2\rangle+e^{i\phi} |V_1,V_2\rangle)$. Since the
output state of the cavity $|HV\rangle$ is already a NOON state in
the circular basis $|HV\rangle=i/\sqrt{2}(|L_1,L_2\rangle+
|R_1,R_2\rangle)$, this state can be transferred into a NOON state
in the $H$/$V$ basis by sending it through an additional quarter
wave plate at 45 degrees.
\begin{figure}[b]
\centering
\includegraphics[width=8cm]{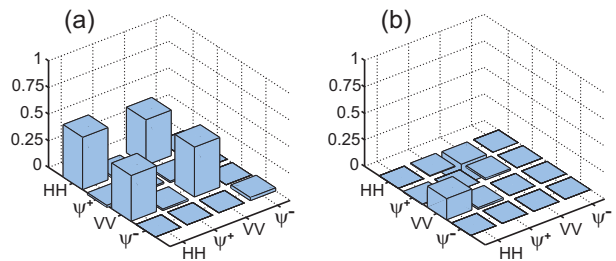}
\caption{(a) Real and (b) imaginary part of the polarization
density matrix of the pair-photon state transformed to a
two-photon NOON state.\label{img:NOON_states}}
\end{figure}
In Figure~\ref{img:NOON_states} real and imaginary parts of the
reconstructed density matrix of a NOON state are displayed. The
coherence of the state is partly imaginary leading to a phase of
$\phi=0.20$ between $HH$ and $VV$ components
(Figure~\ref{img:NOON_states}(b)), which is however of no
importance in the following. The fidelity of this state with the
corresponding ideal two-photon NOON state
$1/\sqrt{2}(|H_1,H_2\rangle+e^{i\phi} |V_1,V_2\rangle)$ is 99\%,
making the state suitable for applications such as
phase-estimation \cite{Mitchell2004}. To demonstrate this ability,
we performed a super-resolving phase experiment. After passing the
NOON state, for this experiment in the circular basis (without
QWP1 and QWP2), through the HWP, the coincidence counts between
the output ports of PBS2 for different HWP settings were recorded.
In Figure~\ref{img:Super-Resolution} the interference fringes of
the coincidences are displayed together with single counts from a
measurement in which one polarization of the cavity output was
blocked.
The period of the coincidence counts oscillations is shorter by a
factor of two compared to the singles, as expected for a
two-photon NOON state. The sinusoidal fit function of the
coincidences shows a high visibility of 90\%.
\\
We have used quantum state tomography to analyze the pair-photon
state from a cavity-enhanced down-conversion source in order to
optimize the indistinguishability of the photons. The highly
indistinguishable photons have then been converted to a
high-fidelity polarization NOON state in the $H$/$V$ basis.
Furthermore, a phase super-resolution measurement using NOON
states has been demonstrated showing that these states are
suitable for applications in quantum imaging and atomic
spectroscopy. For the cavity-enhanced down-conversion source used
in this experiment, applications in atomic spectroscopy are
especially promising since the photon wavelength is at a rubidium
resonance. In addition, the photons are spectrally tailored, such
that even after filtering them to a bandwidth of a few MHz, which
is comparable to the natural linewidth of atomic rubidium, a high
count rate of 70~pairs/(s~mW~MHz) is expected \cite{Cere2009,
Wolfgramm2008}.
\\
\begin{acknowledgments}
We thank R.~B.~A.~Adamson and R.~L.~Kosut for help on the
reconstruction code as well as X.~Xing, A.~M.~Steinberg and
A.~Predojevi\'{c}, who participated in early work for this
experiment. In addition we want to thank Y.~de~Icaza~Astiz for
useful discussions.
\\
This work was supported by the Spanish Ministry of Science and
Innovation under the Consolider-Ingenio 2010 Project ``Quantum
Optical Information Technologies'' and the ILUMA project (Ref.
FIS2008-01051) and by an ICFO-OCE collaborative research program.
F.~W. is supported by the Commission for Universities and Research
of the Department of Innovation, Universities and Enterprises of
the Catalan Government and the European Social Fund.
\end{acknowledgments}

\end{document}